\documentstyle[12pt,equations]{article}
\begin{document}
\newcommand{\dkl}{\delta \kappa_\Lambda}
\newcommand{\kx}{\kappa}
\newcommand{\sx}{\sigma}
\newcommand{\lx}{\lambda}
\newcommand{\Lx}{\Lambda}
\newcommand{\rht}{\tilde{\rho}}
\newcommand{\rhz}{\rho_0}
\newcommand{\be}{\begin{equation}}
\newcommand{\ee}{\end{equation}}
\newcommand{\een}{\end{subequations}}
\newcommand{\ben}{\begin{subequations}}
\newcommand{\beq}{\begin{eqalignno}}
\newcommand{\eeq}{\end{eqalignno}}
\newcommand{\lsim}{\begin{array}{c}<\vspace{-0.32cm}\\\sim\end{array}}
\newcommand{\gsim}{\begin{array}{c}>\vspace{-0.32cm}\\ \sim\end{array}}
\pagestyle{empty}
\noindent
OUTP 94-17 P \\
HD-THEP-94-29
\vspace{3cm}
\begin{center}
{\bf \Large Analytical Solutions of Exact}
\\ \medskip
{\bf \Large Renormalization Group Equations}
\\ \vspace{1cm}
N.~Tetradis$^{\rm a,}$\footnote{E-mail: TETRADIS@THPHYS.OX.AC.UK}
and
D.F.~Litim$^{\rm b,}$\footnote{
E-mail: CU9@IX.URZ.UNI-HEIDELBERG.DE \\
${}\,\,\, \quad $ Address after 30${}^{th}$ of September 1995: \\
${}\,\,\, \quad$  Imperial College,
Blackett Laboratory, Prince Consort Road, London SW7 2BZ, U.K.}
\\ \vspace{1cm}
{}$^{\rm a}$Theoretical Physics\\ University of Oxford\\
1 Keble Road\\ Oxford OX1 3NP, U.K.\\[2ex]
{}$^{\rm b}$Institut  f\"ur Theoretische Physik\\ Universit\"at Heidelberg\\
Philosophenweg 16\\69120 Heidelberg, Germany\\
\vspace{1cm}
\abstract{
We study exact renormalization group equations in the
framework of the effective average action. We present
analytical solutions for the scale dependence of
the potential in a variety of models.
These solutions display a rich spectrum of physical
behaviour such as fixed points governing the
universal behaviour near second order phase transitions,
critical exponents,  first order
transitions (some of which are radiatively induced) and
tricritical behaviour.
}
\end{center}

\clearpage

\setlength{\baselineskip}{15pt}
\setlength{\textwidth}{16cm}
\pagestyle{plain}
\setcounter{page}{1}

\newpage

\setcounter{equation}{0}
\renewcommand{\theequation}{{\bf 1.}\arabic{equation}}

\section*{1.
Introduction}

The solution of an exact renormalization group equation
\cite{reneq1}-\cite{ellwanger}
is a particularly difficult task.
The reason is that such an equation
describes the scale dependence of
an effective action, which is characterized
by infinitely many couplings multiplying the
invariants consistent with the symmetries of the model
under consideration.
As a result an exact renormalization group equation
corresponds to infinitely many evolution equations for the
couplings of the theory.
The crucial step is
developing efficient approximation schemes which can
reduce the complexity of the problem while
capturing the essential aspects of the physical system.
Perturbative expansions have been used for proofs
of perturbative renormalizability
\cite{polchinski,pertren}, while the
powerful $\epsilon$-expansion \cite{reneq1,epsilon}
has been employed for the study of fixed points governing
second order phase transitions in three dimensions.
In many cases
evolution equations for truncated
forms of the effective action have been solved through
a combination of analytical and numerical
methods (see the review \cite{review} and references therein,
\cite{hasenfratz,bound}, \cite{vv}-\cite{num}.
Also, numerical solutions for
the fixed point potential of three-dimensional scalar
theories have been computed in ref.\ \cite{numer,morris}.

In this paper we present analytical solutions
of equations which describe the evolution of the running
effective potential.
We are interested in describing the evolution not only
near the possible fixed points, but also far away from them.
This allows us to study the influence of the initial conditions
(the parameters of the classical theory)
on the renormalized theory. We can also investigate theories
with first order phase transitions, which do not exhibit universal
behaviour. The biggest part of this work
concerns the $O(N)$-symmetric scalar theory in the large $N$ limit,
the spherical model \cite{spherical}.
We obtain analytical solutions of the
evolution equation which describes
the dependence of the running
effective potential $U_k$ on an effective infrared cutoff $k$.
This equation, and our solutions, are exact in the large $N$ limit.
In three dimensions,
we show that for a certain range of classical parameters
the evolution of the critical theory
leads to an infrared attractive fixed point
(the Wilson-Fisher fixed point). This results in
a second order phase transition with universal behaviour.
For a different parameter range the theory has a first order phase
transition. A tricritical line separates the two regimes.
In four dimensions we describe how the renormalized quartic
coupling evolves to zero as the infrared cutoff is removed.
Finally in two dimensions we explicitly demonstrate the absence
of symmetry breaking and a phase transition for this model.
The incorporation of physical phenomena of such a wide
range is the most prominent and promising characteristic of
our analytical solutions.
We subsequently study an approximate evolution equation for the
running effective potential in
the context of the Abelian Higgs model in four dimensions.
We demonstrate how its solution predicts a first order phase transition
driven by radiative corrections (the Coleman-Weinberg mechanism).
Parts of our solutions for the $O(N)$-symmetric scalar theory
have been obtained previously \cite{vv,review}.
However, the starting point for our derivation is
a recently proposed renormalization group equation
\cite{exeq}, and our results demonstrate the
complete agreement of its physical predictions
with those of other
renormalization group equations \cite{reneqex,nicoll,hasenfratz}.
We also give a detailed discussion of the
full range of the running effective potential (including
its non-convex part), and consider a variety of
regions for the classical
parameters of the theory.

We work in the framework of the effective average action
$\Gamma_k$ \cite{exeq,averact}, which has been used for the
study of second and first order
phase transitions for a variety of four-dimensional
field theories at non-zero temperature
\cite{trans}-\cite{abel}.
Also, in ref.\ \cite{indices} an approximate solution
of the evolution equation for the potential
for the three-dimensional $O(N)$-symmetric scalar theory
in the large $N$ limit has been given.
The effective average action
$\Gamma_k$ results from the integration
of quantum fluctuations with characteristic momenta
$q^2 \geq k^2$. It
interpolates between the classical action $S$
for $k$ equal to the
ultraviolet cutoff $\Lx$
of the theory (no integration of modes)
and the effective action $\Gamma$ for $k=0$
(all the modes are integrated). Its dependence on $k$ is given
by an exact renormalization group equation with the typical form
($t=\ln (k/\Lx$))
\be
\frac{\partial}{\partial t} \Gamma_k
= \frac{1}{2} {\rm Tr} \bigl\{ (\Gamma_k^{(2)} + R_k)^{-1}
\frac{\partial}{\partial t} R_k \bigr\}.
 \label{oneone} \ee
Here $\Gamma_k^{(2)}$ is the second functional
derivative with respect to the fields,
$R_k$ is the effective infrared cutoff which prevents the
integration of modes with $q^2 \leq k^2$,
and the trace implies integration over all Fourier modes of the
fields.
We work with an approximation which neglects the
effects of wavefunction renormalization.
Therefore,
only a classical kinetic term in the effective average
action is kept, which takes, for an $O(N)$-symmetric
scalar theory, the form
\be
\Gamma_k =
\int d^dx \bigl\{ U_k(\phi)
+ \frac{1}{2} \partial^{\mu} \phi_a
\partial_{\mu} \phi^a \bigr\},
\label{twoeight} \ee
and all invariants which involve more derivatives
of the fields are neglected.
With this approximation, eq.~(\ref{oneone}) can be
turned into an evolution equation for the potential $U_k$.
First we shall discuss the $O(N)$-symmetric scalar theory
in the large $N$ limit. In
three dimensions the fixed point solution which
governs the second order phase transition will be identified. We shall also
show that, for a certain parameter range, the theory
has a first order phase transition. The two regions in
parameter space are separated by a tricritical line. We shall
also study the same theory in four and two dimensions.
Finally we shall discuss the Abelian Higgs model in four dimensions,
for which the radiatively induced first order transition will be
reproduced.

\setcounter{equation}{0}
\renewcommand{\theequation}{{\bf 2.}\arabic{equation}}

\section*{2.
The $O(N)$-symmetric scalar theory}

We first consider the evolution equation describing the
dependence of the effective average potential $U_k$ on the scale
$k$ in arbitrary dimensions $d$, for an $O(N)$-symmetric
scalar theory. As we have mentioned in the introduction, we
neglect the effects of wavefunction renormalization.
The evolution equation reads ($t =\ln (k/\Lx)$)
\cite{exeq}
\be
\frac{\partial U_k(\rho)}{\partial t}  = v_d\;
\int_0^{\infty}
dx\; x^{\frac{d}{2}-1} \; \frac{\partial P}{\partial t}
\left\{
\frac{N-1}{P+U'_k(\rho)}
+ \frac{1}{P+U'_k(\rho)+ 2U''_k(\rho)\rho}
\right\}.
\label{one}
\ee
Here $\rho=\frac{1}{2}\phi^a \phi_a,~a=1...N,$
and primes denote derivatives with respect to $\rho$.
The variable $x$ denotes momentum squared $x=q^2$,
and
\be
v_d^{-1} = 2^{d+1} \pi^\frac{d}{2} \Gamma \left( \frac{d}{2}
\right).
\label{extra} \ee
The inverse average propagator
\be
P(x) = \frac{x}{1- f^2_k(x)}
\label{two}
\ee
incorporates an effective infrared cutoff $k$, such that modes
with $x=q^2 \ll k^2$ do not propagate. The ``cutoff'' function
$f^2_k(x)$ is given by
\be
f^2_k(x) = \exp \biggl\{
-2 a \left( \frac{x}{k^2} \right)^b \biggr\}
\label{twothree}
\ee
and can be made sharper or smoother by an appropriate choice of the
two free parameters $a,b$.
Up to effects
from the wavefunction renormalization
eq.~(\ref{one}) is an exact
non-perturbative evolution equation \cite{exeq}.
It is easy to recognize the first term in the r.h.s. of
eq.~(\ref{one}) as the contribution of the $N-1$ Goldstone modes
($U'_k$ vanishes at the minimum).
The second term is related to the radial mode.
After performing the momentum integration the evolution equation
(\ref{one})
becomes a partial differential equation for $U_k$ with
independent variables $\rho$ and $t$.
The effective average potential interpolates
between the classical potential $V$
for $k=\Lx$ (with $\Lx$ the
ultraviolet cutoff) and the effective potential $U$ for $k=0$
\cite{exeq}. As a result
the solution of eq.~(\ref{one}) with the initial
condition $U_{\Lx}(\rho)=V(\rho)$
uniquely determines,
for $k \rightarrow 0$, the
effective 1PI vertices at zero momentum for the renormalized theory.
In order to write eq.~(\ref{one}) in a scale invariant form
it is convenient to define the variables
\beq
\rht = &k^{2-d} \rho \nonumber \\
u_k(\rht) = &k^{-d} U_k(\rho).
\label{three} \eeq
In terms of these eq.~(\ref{one})
can be written as (for the details see
refs.\ \cite{exeq,averact,trans,indices})
\be
\frac{\partial u'}{\partial t} =~ -2 u' +(d-2)\rht u''
+ v_d (N-1) u'' L^d_1(u')
+ v_d (3 u'' + 2 \rht u''') L^d_1(u'+2 \rht u'').
\label{four} \ee
Primes on $u$ denote derivatives with respect to $\rht$ and we
omit the subscript $k$ in $u_k$ from now on.
The functions $L^d_1(w)$ are given by
\be
L^d_1(w) = -2  (2 a)^{\frac{2-d}{2 b}}
\int_0^{\infty} dy~y^{\frac{d-2}{2 b}} e^{-y}
\left[ 1 + \left( \frac{2 a}{y} \right)^{\frac{1}{b}}
\left(1 - e^{-y} \right) w \right]^{-2}
\label{five} \ee
(where we have used the variable
$y = 2 a \left( x/k^2 \right)^{b}$).
They introduce ``threshold'' behaviour
in the evolution equation, which results in the decoupling
of heavy modes. They approach a constant for vanishing argument,
and vanish for large arguments.
For general $a$ and $b$ they do not have a
simple analytical form. However, in the sharp cutoff limit
$b \rightarrow \infty$
they are given by the simple expression
\be
L^d_1(w) =~-\frac{2}{1+ w}.
\label{six}  \ee
The resulting evolution equation (with an appropriate rescaling
of the fields) is identical to the one introduced in
ref.\ \cite{reneqex} and studied in
refs.\ \cite{reneqex,hasenfratz,morris}.
Also, in the large $N$ limit (see below) it is the same as the
evolution equation \cite{review} which results from the exact
renormalization group equation of ref. \cite{nicoll},
for which solutions are given in ref. \cite{vv}.
Our derivation establishes the agreement of
the physical predictions resulting from the formalism of
the effective average action with those of previous renormalization
group formulations.
We should point out that the use of a sharp cutoff
renders the effective average action
$\Gamma_k$ non-local. This complication, however, does
not affect our analysis, as we have neglected the
effect of the higher derivative terms on the evolution
equation for the potential. For studies which take into
account the effects of the wavefunction renormalization
\cite{indices} the use of a smooth cutoff is
necessary.

Even with the simple form of eq.~(\ref{six}) for the
``threshold'' functions, the evolution equation
(\ref{four}) remains a non-linear partial differential
equation which seems very difficult to solve exactly.
\footnote{See ref. \cite{num} for a numerical solution of
eq. (\ref{four}).}
The problems arise from
the contribution from the radial mode
to the r.h.s. of eq.~(\ref{four}) (the last term).
An enormous simplification is achieved, however,
if this contribution is neglected.
This is justified in the large $N$ limit, in which
the contributions of the Goldstone modes dominate
the evolution, and the contribution of the radial mode
becomes a subleading effect. Another important simplification
results from the fact that the anomalous dimension
of the field is zero to leading order in $1/N$, in
four and three dimensions which are of most interest \cite{zinn}.
As a result, our approximation of neglecting the
wavefunction renormalization is justified in this limit
and our solution of eq.~(\ref{four}) becomes exact.
We should point out that the contributions of the radial
mode cannot be neglected even in the large $N$ limit in
some cases. For certain forms of the classical
potential, the approach to convexity for the effective
potential depends crucially
on the radial mode. An example of this behaviour will be
given in the following.
For $N \gg 1$
the simplified evolution equation reads
\be
\frac{\partial u'}{\partial t}
- (d-2) \rht \frac{\partial u'}{ \partial \rht}
+ \frac{NC}{1+u'}
\frac{\partial u'}{ \partial \rht}
+ 2 u' = 0,
\label{seven} \ee
with $C = 2 v_d$.
It is first order in both independent variables and
can be solved with the method of characteristics.

\subsection*{
a) Three dimensions}
We are interested in the behaviour of the theory
in three dimensions, where a non-trivial fixed point
structure arises. The most
general solution of the partial differential equation
(\ref{seven}) for $d=3$ is given by the relations
\beq
&\frac{\rht}{\sqrt{u'}} - \frac{NC}{\sqrt{u'}}
+ NC \arctan \left( \frac{1}{\sqrt{u'}} \right)
= F \left( u' e^{2t} \right)~~~~~~~~~~~~~~~{\rm for}~~u'>0
\label{eight} \\
&\frac{\rht}{\sqrt{-u'}} - \frac{NC}{\sqrt{-u'}}
- \frac{1}{2} NC
\ln \left( \frac{1-\sqrt{-u'}}{1+\sqrt{-u'}} \right)
= F \left( u' e^{2t} \right)~~~~~~~{\rm for}~~u'<0,
\label{nine} \eeq
with
\be
C = 2 v_3 = \frac{1}{4 \pi^2}.
\label{trala} \ee
The function $F$ is undetermined until initial conditions are specified.
For $t=0$ ($k=\Lambda$), $U_k$ coincides with the classical
potential $V$.
The initial condition, therefore, reads
\be
u'(\rht,t=0) = \Lx^{-2} V'(\rho).
\label{ten} \ee
This uniquely specifies $F$ and we obtain
\beq
\frac{\rht}{\sqrt{u'}} - &\frac{NC}{\sqrt{u'}}
+ NC \arctan \left( \frac{1}{\sqrt{u'}} \right)
= \nonumber \\
&\frac{G\left( u' e^{2t} \right)}{\sqrt{u'} e^t}
- \frac{NC}{\sqrt{u'} e^t}
+ NC \arctan \left( \frac{1}{\sqrt{u'}e^t} \right)
{}~~~~~~~~~~{\rm for}~~u'>0
\label{eleven} \\
\frac{\rht}{\sqrt{-u'}} - &\frac{NC}{\sqrt{-u'}}
- \frac{1}{2} NC
\ln \left( \frac{1-\sqrt{-u'}}{1+\sqrt{-u'}} \right)
= \nonumber \\
&\frac{G\left( u' e^{2t} \right)}{\sqrt{-u'} e^t}
- \frac{NC}{\sqrt{-u'}e^t}
- \frac{1}{2} NC
\ln \left( \frac{1-\sqrt{-u'} e^t}{1+\sqrt{-u'} e^t} \right)
{}~~~~~{\rm for}~~u'<0,
\nonumber \\
{}~~&~~
\label{twelve} \eeq
with the function $G$
determined by inverting eq.~(\ref{ten})
and solving for $\rht$ in terms of $u'$
\be
G(u') = \rht(u')|_{t=0}.
\label{thirteen} \ee
We are also interested in the scale invariant (independent of $t$)
solutions of eq.\ (\ref{seven}).
These solutions correspond to the possible fixed points
of the evolution and are obtained by setting the first
term in the l.h.s. of eq.\ (\ref{seven}) equal to zero.
We can easily identify the ``trivial'' solution $u'_\star=0$,
which is independent of the number of dimensions.
For $d=3$
the only other solution which remains finite for finite $\rht$
is given by the relations
\beq
&\frac{\rht}{\sqrt{u'_\star}} - \frac{NC}{\sqrt{u'_\star}}
+ NC \arctan \left( \frac{1}{\sqrt{u'_\star}} \right)
= \frac{\pi}{2} NC~~~~~~~~~~~~~~~{\rm for}~~u'_\star>0
\label{eighteen} \\
&\frac{\rht}{\sqrt{-u'_\star}} - \frac{NC}{\sqrt{-u'_\star}}
- \frac{1}{2} NC
\ln \left( \frac{1-\sqrt{-u'_\star}}{1+\sqrt{-u'_\star}} \right)
= 0~~~~~~~~~~~~~{\rm for}~~u'_\star<0.
\label{nineteen} \eeq
This solution corresponds to the Wilson-Fisher fixed point.
A large part of the above results has been obtained in the past
\cite{vv,review}. However, we have explicitly presented here the
solution for the non-convex part of the potential. We are also
mainly interested in studying the
various phase transitions which result for
different ranges of classical parameters of the theory.

I) {\it Classical $\phi^4$ theory:}
Let us first consider a theory with a
quartic classical potential.
The initial condition can be written as
\be
u'(\rht,t=0) = \lx_{\Lambda} (\rht - \kappa_\Lx),
\label{fourteen} \ee
with
\be
\kappa_{\Lambda} = \frac{\rho_{0 \Lx}}{\Lx},~~~~~\lx_{\Lx} =
\frac{{\bar \lx}_{\Lx}}{\Lambda}
\label{fifteen} \ee
the rescaled (dimensionless)
minimum of the potential and quartic coupling
respectively.
The function $G$
in eqs.\ (\ref{eleven}), (\ref{twelve})
is now given by
\be
G(x)= \kappa_{\Lx} + \frac{x}{\lx_{\Lx}}.
\label{sixteen} \ee
The typical form of the effective average potential
$U_k(\rho)$ at
different scales $k$, as given by eqs.\ (\ref{eleven}),
(\ref{twelve}), is presented in fig.~1.
The theory at the ultraviolet cutoff is defined in the
regime with spontaneous symmetry breaking,
with the minimum of the potential at
$\rho_{0 \Lx} = \kappa_\Lx \Lx \not= 0$.
At lower scales $k$ the minimum of the potential
moves continuously closer to zero, with no secondary minimum
ever developing. We expect a second order
phase transition (in dependence to $\kappa_\Lx$) for
the renormalized theory at $k=0$.
There is a critical value for the minimum of the classical potential
\be
\kx_\Lambda = \kx_{cr} = NC,
\label{seventeen} \ee
for which the scale invariant (fixed point)
solution of eqs.\ (\ref{eighteen}), (\ref{nineteen})
is approached in the limit
$t \rightarrow -\infty$ ($k \rightarrow 0$).
Eqs.\ (\ref{eighteen}), (\ref{nineteen})
describe a potential $u$ which has a minimum at a constant value
\be
\kx(k) = \kx_\star = NC.
\label{twenty} \ee
This leads to
a potential $U_k(\rho)$ with a minimum at
$\rho_0(k) = k \kappa_\star  \rightarrow 0$
for $k \rightarrow 0$, which corresponds to the phase transition
between the spontaneously broken and the symmetric phase.
(The values for $\kx_{cr}$ and $\kx_{\star}$ coincide, but this is accidental.)
For the second and third $\rht$-derivative of $u$ at the minimum
$\lx = u''(\kx)$,
$\sx = u'''(\kx)$
we find
\beq
\lx(k) =& \lx_\star = \frac{1}{NC}
\label{twentyone} \\
\sx(k) =& \sx_\star = \frac{2}{3 (NC)^2},
\label{extranu} \eeq
and similar fixed point values for the higher derivarives of $u$.
For $1 \ll \rht/ NC \ll (\pi/2) e^{-t} $ the rescaled potential
$u$ has the form
\be
u'_\star(\rht) = \left( \frac{2}{\pi  NC} \right)^2 \rht^2.
\label{twentytwo} \ee
Notice that the region of validity of eq.~(\ref{twentytwo}) extends to
infinite $\rht$ for $t \rightarrow - \infty$.
{}From eq.~(\ref{twentytwo}) with $t \rightarrow -\infty
(k \rightarrow 0)$ we obtain for the effective potential at the
phase transition
\be
U_\star(\rho) = \frac{1}{3} \left( \frac{2}{\pi NC} \right)^2 \rho^3.
\label{twentythree} \ee

Through eqs.\ (\ref{eleven}), (\ref{twelve})
we can also study solutions which deviate slightly from the
scale invariant one.
For this purpose we define a classical potential
with a minimum
\be
\kx_\Lambda = \kx_{cr} +  \dkl,
\label{twentyfour} \ee
with $|\dkl| \ll NC$.
A typical solution is depicted in fig.~2.
The potential starts its evolution with its classical
form at $k=\Lx$. It subsequently evolves
towards the scale invariant solution
given by eqs.~(\ref{eighteen}), (\ref{nineteen}).
It stays very close this solution for a long ``time''
$t$, which can be rendered arbitrarily long
for sufficiently small $|\dkl|$.
Eventually it deviates towards the phase with spontaneous symmetry
breaking or the symmetric one. In fig.~2 $\dkl$ was chosen negative,
so that the evolution for $k \rightarrow 0$ leads to the symmetric
phase. The curvature of the potential at the origin becomes
positive and $u'(0)$ diverges so that the renormalized mass term
$U'_k(0) = k^2 u'(0)$ reaches a constant value.
The evolution of the minimum of the potential is given by
\be
\kx(k) = \kx_\star +  \dkl e^{-t}.
\label{twentyfive} \ee
Also for $\lx$ we find
\be
\lx(k) = \frac{\lx_\star}{1 + \left( \frac{\lx_\star}{\lx_{\Lambda}}
-1 \right)
e^t }.
\label{twentysix} \ee
Eq.~(\ref{twentyfive}) indicates that the minimum of $u$ stays
close to the fixed point value $\kx_\star$ given by
eq.~(\ref{twenty}), for a very long ``time''
$|t| < -\ln |\dkl|$.
For $|t| > -\ln |\dkl|$ it deviates from the fixed point,
either towards the phase with spontaneous symmetry breaking
(for $\dkl > 0$), or the symmetric one
(for $\dkl < 0$).
Eq.~(\ref{twentysix}) implies an attractive fixed point for $\lx$,
with a value given by eq.~(\ref{twentyone}).
Similarly the higher derivatives are attracted to their fixed point
values.
The full phase diagram corresponds to a second order phase transition.
For $\dkl > 0$ the system ends up in the phase with spontaneous
symmetry breaking, with
\be
\rhz = \lim_{k\rightarrow 0} \rhz(k) = \lim_{k \rightarrow 0}
k \kappa(k) =  \dkl \Lx.
\label{twentyseven} \ee
In this phase the renormalized
quartic coupling
approaches zero linearly with $k$
\be
\lx_R = \lim_{k \rightarrow 0} k \lx(k) =
\lim_{k \rightarrow 0}k \lx_\star =0.
\label{twentyeight} \ee
The fluctuations of the Goldstone bosons
lead to an infrared free theory in the phase with
spontaneous
symmetry breaking.
For $\dkl < 0$, $\kx(k)$ becomes zero at a scale
\be
t_s = - \ln \left( \frac{\kx_\star}{|\dkl|} \right)
\label{twentynine} \ee
and the system ends up in the symmetric regime ($\rhz=0$).
{}From eq. (\ref{eleven}), in the limit
$t \rightarrow -\infty$, with $u',u'',u''' \rightarrow \infty$,
so that
$u'e^{2t} \sim |\dkl|^2$, $u''e^{t} \sim |\dkl|$, $u''' \sim 1$,
we find
\be
U(\rho) = U_0(\rho) = \left( \frac{2}{ \pi NC} \right)^2
\left[ |\dkl|^2 \Lx^2 \rho + |\dkl| \Lx \rho^2 +
\frac{1}{3} \rho^3 \right].
\label{thirty} \ee
Notice how every reference to the classical theory has
disappeared in the above expression. The effective
potential of the critical theory is determined uniquely in
terms of $\dkl$, which measures the distance from the
phase transition.
The above results are in exact agreement with
refs.~\cite{spherical,vv,review,zinn,indices,largen}.
In particular the values
for the critical exponents $\beta$, $\nu$, describing
the behaviour of the system very close to the phase transition,
correspond to the large $N$ limit of the model (for details
see section 10 of ref.\ \cite{indices})
\beq
\beta =& \lim_{\dkl \rightarrow 0^+}
\frac{d \left( \ln \sqrt{\rhz} \right)}{d (\ln \dkl)} = 0.5
\nonumber \\
\nu =& \lim_{\dkl \rightarrow 0^-}
\frac{d \left( \ln m_R \right)}{d (\ln  |\dkl| )} =
\frac{d \left( \ln \sqrt{U'(0)} \right)}{d (\ln |\dkl| )}
= 1.
\label{exxt} \eeq
We also point out that, for a theory with spontaneous
symmetry breaking , we can use eq.~(\ref{twelve})
in order to study the ``inner'' part of the potential.
In particular, for $\rht = 0$ and $t \rightarrow -\infty$
eq.~(\ref{twelve}) predicts a potential $u$ which asymptotically behaves
as
\be
\lim_{t \rightarrow -\infty} u'(0) = -1.
\label{thirtyone} \ee
This leads to an effective average potential $U_k$ which becomes convex with
\be
\lim_{k \rightarrow 0} U'_k(0) = - k^2,
\label{thirtytwo} \ee
in agreement with the detailed study of ref.\ \cite{convex}.
\par
II) {\it Classical $\phi^6$ theory:}
As a second example we consider a theory
defined through a classical potential with a $\rho^3$ ($\phi^6$)
term
\be
u'(\rht,t=0) = \lx_{\Lambda} (\rht - \kappa_\Lx)
+ \frac{\sx_{\Lambda}}{2} (\rht - \kappa_\Lx)^2,
\label{fifty} \ee
where $\kappa_\Lx$, $\lx_\Lx$ are defined in eq.~(\ref{fifteen})
and the coupling $\sx_\Lx$ is dimensionless in $d=3$.
The function $G$ in eqs.\ (\ref{eleven}), (\ref{twelve})
is now given by
\beq
G(x) =& \kappa_\Lx + \frac{-\lx_\Lx
+ \sqrt{\lx^2_\Lx + 2 \sx_\Lx x}}{\sx_\Lx}
{}~~~~~~~~~~{\rm for}~~u''>0
\label{fiftyone} \\
G(x) =& \kappa_\Lx + \frac{-\lx_\Lx
- \sqrt{\lx^2_\Lx + 2 \sx_\Lx x}}{\sx_\Lx}
{}~~~~~~~~~~{\rm for}~~u''<0.
\label{fiftytwo} \eeq
We distinguish two regions in parameter
space which
result in two different types of
behaviour for the theory: \\
(a) For $\kx_\Lx < 2 \lx_\Lx/\sx_\Lx$
the classical potential has only one minimum at
$\rho_{0 \Lx} = \kx_\Lx \Lx$. Near this minimum the
initial condition of eq.~(\ref{fifty}) is
very well approximated by eq.~(\ref{fourteen}). As a result,
for $\kx_\Lx$ near the critical value
of eq.~(\ref{seventeen}), the critical theory has exactly the
same behaviour as for a quartic classical potential.
The running potential first approaches the fixed point solution
of eqs.\ (\ref{eighteen}), (\ref{nineteen})
(notice that $\kx_\star <2  \lx_\star/\sx_\star$), and subsequently evolves
towards the phase with spontaneous symmetry breaking or the
symmetric one.
The behaviour of the critical
theory for $k =0$ is determined only by the distance from
the phase transition (as measured by $\dkl$), without any
memory of the details of the classical theory.
This is a manifestation of universality, typical of second order
phase transitions.
For $\kx_\Lx > 2 \lx_\Lx/\sx_\Lx$
the classical potential has two minima, one at
 the origin and one
at $\rho_{0 \Lx} = \kx_\Lx \Lx$. The minimum at the origin
is less deep
for $\kx_\Lx < 3 \lx_\Lx/\sx_\Lx$.
Again, for
$\kx_\Lx$ near $\kx_{cr}$ the scale invariant solution is approached.
We demonstrate this type of behaviour in fig.~3 for a theory with
$\kx_\Lx = \kx_{cr}$. The classical potential includes a
$\phi^6$ term and has a (less deep) second minimum at the origin,
but the universal scale invariant solution
is again
approached for $k \rightarrow 0$. Small deviations of $\kx_\Lx$
from the critical value result in universal behaviour for the
renormalized theory.
\\
(b)
The minimum of the classical potential at the origin is deeper
for $\kx_\Lx > 3 \lx_\Lx/\sx_\Lx$.
An example of the evolution of the effective average potential
$U_k(\rho)$
for such a theory is given in fig.~4. The minimum of the
potential at non-zero $\rho$ moves towards the origin
for decreasing scale $k$. In the same time the positive
curvature at the origin decreases. The combined effect
is (very crudely) similar to the whole potential
being shifted to the left of the graph.
As a result the minimum at the origin becomes shallower.
For a certain range of the parameter space (for small enough
$\kx_\Lx$, such as chosen for fig.~4) the minimum away from
the origin becomes the absolute minimum of the potential at some point
during the evolution. This results in a discontinuity in the
running order parameter.
Finally the absolute minimum of the potential
settles down at some non-zero $\rhz$.
For even larger $\kx_\Lx$ the minimum at the origin is
deep enough for the evolution to stop while this minimum is still
the absolute minimum of the potential.
When the minimum of the renormalized potential $\rhz$
(which is obtained at the end of the evolution)
is considered as a function of $\kx_\Lx$, a discontinuity is
observed in the function $\rhz(\kx_\Lx)$. This indicates a first
order phase transition.
Unfortunately, an exact quantitative
determination of the region in
parameter space which results in first order transitions is not
possible within the approximations we have used.
The reason for this is the omission of
the ``threshold'' function for the radial mode which includes
the term $2 \rht u''$.
As a result our approximation is not adequate for
dealing with the shape of the barrier
in the limit $k \rightarrow 0$,
where the theshold function
for the radial mode becomes important.
Also the approach to
convexity cannot be reliably discussed
(in contrast to
the case of a classical $\phi^4$ potential).
If the shape of the barrier cannot be reliably determined
the relative depth of the two minima cannot be calculated, and
our discussion is valid only at the qualitative level.
\par
However, more information can be extracted from our results.
As long as we concentrate on regions of the potential
away from the top of the barrier the solution given by eqs.\
(\ref{eleven}), (\ref{fiftyone}), (\ref{fiftytwo})
is reliable. This means that we can study the potential
around its two minima.
We are interested in the limit
$t \rightarrow -\infty$
($k \rightarrow 0$), with
$U' = u'e^{2t}$, $\rho =\rht e^{t}$ approaching finite values.
The form of the potential near the
minimum away from the origin is determined by eqs.\
(\ref{eleven}), (\ref{fiftyone}).
We find
\be
\frac{\rho}{\Lx} - \kx_\Lx + NC =
\frac{- \lx_\Lx + \sqrt{\lx_\Lx^2+2\sx_\Lx
\frac{U'}{\Lx^2}}}{\sx_\Lx}
+ NC \sqrt{\frac{U'}{\Lx^2}}
\arctan \left( \frac{1}{\sqrt{\frac{U'}{\Lx^2}}} \right).
\label{seventyone} \ee
The minimum $\rhz$ (where $U'(\rhz)=0$) is located at
$\rhz = (\kx_\Lx - NC) \Lx = \dkl \Lx $. This
requires $\dkl \geq 0$.
Eqs.\ (\ref{eleven}), (\ref{fiftytwo})
describe the
form of the potential around the minimum at the origin.
Similarly as above we find
\be
\frac{\rho}{\Lx} - \kx_\Lx + NC =
\frac{- \lx_\Lx - \sqrt{\lx_\Lx^2+2\sx_\Lx
\frac{U'}{\Lx^2}}}{\sx_\Lx}
+ NC \sqrt{\frac{U'}{\Lx^2}}
\arctan \left( \frac{1}{\sqrt{\frac{U'}{\Lx^2}}} \right).
\label{seventytwo} \ee
In the parameter range
$\kx_\Lx - 2 \lx_\Lx/\sx_\Lx,
2 \lx_\Lx/\sx_\Lx \gg NC$ the above
solution reproduces the classical potential, with
a large positive curvature $U'(0)/\Lx^2$ at the origin.
This is due to the fact that the fluctuations
which renormalize the potential around the
origin are massive, with their
masses acting as an effective infrared cutoff.
For the above parameter range these masses
are of the order of the ultraviolet cutoff $\Lx$ and
no renormalization of the potential takes place.
This is in contrast with the form of the potential
near the minimum $\rhz$ away from the origin. The presence
of the Goldstone modes in this region always results in
strong renormalization.
There is a range of parameters for which
the curvature at the origin becomes zero.
It is given by the relation
\be
\kx_\Lx = NC + 2 \lx_\Lx/\sx_\Lx.
\label{seventythree} \ee
For this range the minimum
at the origin disappears and the potential has only one minimum
at $\rhz  = \dkl \Lx = (\kx_\Lx - NC) \Lx = 2 \lx_\Lx/\sx_\Lx $.
The above condition does not determine precisely
the first order phase transition, as this takes place when
the two minima are degenerate, and
not when the minimum at the origin disappears.
However, it provides a good estimate of its
location. The discontinuity in the order parameter
is expected to be ${\cal{O}}(\dkl)$.
Weakly first order transitions are obtained for
$\lx_\Lx \rightarrow 0$.
We should emphasize that eq.~(\ref{seventytwo})
is not valid for arbitrarily small $U'/\Lx^2$.
This would correspond to a range of the potential
near the top of the disappearing barrier, where
we know that our approximation fails.
This is another reason why
eq.~(\ref{seventythree}) is only indicative
of the location of the first order phase transition.
\par
We have identified two critical surfaces in parameter space.
We saw in (a)
that the surface $\kx_\Lx = NC$ corresponds to second order
phase transitions. Also in (b) we argued that the
surface $\kx_\Lx = NC + 2 \lx_\Lx/\sx_\Lx$
corresponds to first order transitions.
As a result we expect tricritical behaviour to
characterize their intersection, which is
given by the line $\kx_\Lx = NC$, $\lx_\Lx = 0$.
This is confirmed if we approach this line close to the
critical surface $\kx_\Lx = NC$.
More specifically we consider a theory with
$0 < -\dkl = -\kx_\Lx + NC \ll NC$ and $\lx_\Lx \ll 1/NC$.
For this choice of parameters the renormalized
theory is in the symmetric phase very close to the
second order phase transition. The form of the potential
is given by eq.~(\ref{seventyone})
with $U'/\Lx^2 \ll 1$
\be
\frac{\rho}{\Lx} +|\dkl| = \frac{1}{\lx_\Lx}
\frac{U'}{\Lx^2} +\frac{\pi}{2} NC
\sqrt{\frac{U'}{\Lx^2}}.
\label{seventyfour} \ee
For $|\dkl|/NC \ll \lx_\Lx NC$
the potential has the universal form of
eq.~(\ref{thirty}). The initial point of the
evolution is sufficiently close to the
critical surface for the flows to
approach the Wilson-Fisher critical point
before deviating towards the symmetric phase.
The critical exponent $\nu$ takes the
large $N$ value $\nu=1$ according
to eq.~(\ref{exxt}).
In the opposite limit
$|\dkl|/NC \gg \lx_\Lx NC$
the potential near the origin is
given by
\be
U(\rho) = \lx_\Lx
\left( |\dkl| \Lx^2 \rho + \frac{1}{2} \Lx \rho^2
\right)
\label{seventyfive} \ee
and the exponent $\nu$ takes its mean field value
$\nu = 0.5$.
A continuous transition from one type of behaviour to the
other (a crossover curve) connects the two parameter regions.
Clearly, the line
$\kx_\Lx = NC$, $\lx_\Lx = 0$ gives tricritical
behaviour with mean field exponents, in agreement with
the analysis of ref. \cite{david}.
\footnote{The Bardeen-Moshe-Bander phenomenon \cite{bmb} was
not considered in our discussion.}

\subsection*{
b) Four dimensions}

In the previous subsection we explored the non-trivial fixed
point structure of the three-dimensional scalar theory.
We now look for similar structure in four dimensions.
The most general solution of the partial differential equation
(\ref{seven}) for $d=4$ is given by the relation
\be
\frac{\rht}{u'} - \frac{NC}{2} \frac{1}{u'}
+ \frac{NC}{2} \ln \left( \frac{1+u'}{|u'|} \right)
= F \left( u' e^{2t} \right),
\label{fourone} \ee
where now
\be
C = 2 v_4 = \frac{1}{16 \pi^2}.
\label{fourtwo} \ee
We consider a quartic classical potential according to
eq. (\ref{fourteen}),
with
\be
\kappa_{\Lambda} = \frac{\rho_{0 \Lx}}{\Lx^2},
{}~~~~~\lx_{\Lx} = {\bar \lx}_{\Lx}.
\label{fourthree} \ee
This uniquely specifies $F$ and we obtain
\beq
\frac{\rht}{u'} - \frac{NC}{2} \frac{1}{u'}
&+ \frac{NC}{2} \ln \left( \frac{1+u'}{|u'|} \right)
=
\frac{1}{\lx_\Lx}
+ \left(\kx_\Lx - \frac{NC}{2} \right)
\frac{1}{u' e^{2t}}
+ \frac{NC}{2} \ln \left( \frac{1+u'e^{2t}}{|u'|e^{2t}} \right).
\nonumber \\
{}~&~\label{fourfour} \eeq
There is again a critical value for the minimum of the
classical potential
\be
\kx_\Lx = \kx_{cr} = \frac{NC}{2}
\label{fourfive} \ee
which separates two possible phases. The scale invariant solution
is obtained for the above value of $\kx_\Lx$ in
the limit $t \rightarrow -\infty$.
The resulting solution is the ``trivial'' one
\be
u'_\star = 0.
\label{foursix} \ee
As a result, no interesting universal (independent of the
classical parameters) structure can be obtained for
this theory in four dimensions.

We can also investigate solutions which deviate
slightly from the critical one, by considering a classical
potential with a minimum given by eq. (\ref{twentyfour}).
The renormalized potential in the limit
$t \rightarrow -\infty$, with $u' \rightarrow \infty$
and finite $U'=u'e^{2t}$,
is given by the relation
\be
\frac{U'}{\Lx^2} + \frac{NC}{2} \lx_\Lx \frac{U'}{\Lx^2}
 \ln \left( \frac{1+\frac{U'}{\Lx^2}}{\frac{U'}{\Lx^2}} \right)
= \lx_\Lx \left (\frac{\rho}{\Lx^2} - \dkl \right)
\label{fourseven} \ee
for $U' \geq 0$.
For theories with characteristic
mass scale much smaller than the ultraviolet cutoff
(such as in the symmetric phase in the vicinity of
the phase transition) we have
${U'}/{\Lx^2} \rightarrow 0$.
This gives for the renormalized quartic coupling
\be
U'' = \frac{\lx_\Lx}{1 - \frac{NC}{2} \lx_\Lx
 \ln \left( \frac{U'}{\Lx^2/e} \right)} \rightarrow 0.
\label{foureight} \ee
The same happens near
the minimum of the potential in the phase with
spontaneous symmetry breaking, where no
renormalized mass can be generated for the radial mode.
All this is in agreement with the arguments for the
``triviality'' of the scalar theory in the limit that
the ultraviolet cutoff is removed
(see for example ref. \cite{zinn} and refs. therein).

\subsection*{
c) Two dimensions}

We finally turn to two dimensions, where the most general solution
of eq. (\ref{seven}) is given by
\be
\frac{2}{NC} \rho + \ln \left( \frac{|u'|}{1+u'} \right)
= F \left( u' e^{2t} \right),
\label{atwoone} \ee
with
\be
C = 2 v_2 = \frac{1}{4 \pi}
\label{atwotwo} \ee
and $\rho$ dimensionless.
We consider a quartic classical potential according to
eq. (\ref{fourteen}),
with
\be
\kappa_{\Lambda} = \rho_{0 \Lx},~~~~~\lx_{\Lx} =
\frac{{\bar \lx}_{\Lx}}{\Lambda^2}.
\label{atwothree} \ee
This leads to
\be
u' e^{2t} =
\lx_\Lx \left( \rho -\kx_\Lx \right)
+ \frac{NC}{2} \lx_\Lx
\ln \left( \frac{1+u' e^{2t}}{(1+u') e^{2t}} \right).
\label{atwofour} \ee
For any choice of classical parameters the
minimum of the effective average potential
runs to zero at a scale
\be
t_s = - \frac{\kx_\Lx}{NC}.
\label{atwofive} \ee
At $t=0$ the renormalized potential has a minimum
at zero and is given by the expression
\be
\frac{U'}{\Lx^2} =
\lx_\Lx \left( \rho -\kx_\Lx \right)
- \frac{NC}{2} \lx_\Lx
\ln \left( \frac{\frac{U'}{\Lx^2}}{1+\frac{U'}{\Lx^2}} \right).
\label{atwosix} \ee
The renormalized theory cannot be defined in the phase
with spontaneous symmetry breaking and there is no
phase transition, in agreement with the Mermin-Wagner
theorem \cite{mermin}.

\setcounter{equation}{0}
\renewcommand{\theequation}{{\bf 3.}\arabic{equation}}

\section*{3.
The Abelian Higgs model in four dimensions:}

We now turn to gauge theories, for which exact renormalization
equations have also been obtained \cite{gauge}. As an example
we discuss the Abelian Higgs model
with $N$ real scalars, in four dimensions.
The evolution equation can be written in the form
(for the details see ref.\ \cite{abel})
\be
\frac{\partial u'}{\partial t} =~ -2 u' + 2 \rht u''
+ (N-1) v_4 u'' L^4_1(u')
+ v_4 (3 u'' + 2 \rht u''') L^4_1(u'+2 \rht u'')
+ 6 v_4 e^2 L^4_1(2 e^2 \rht),
\label{eighty} \ee
with $v_4$ given by eq.~(\ref{extra}).
We have again neglected the small wavefunction
renormalization effects for the scalar field. We recognize the
contributions of the Goldstone and radial modes. The last term
in eq.~(\ref{eighty}) is the contribution of the gauge field.
It involves the gauge coupling $e^2$,
whose evolution can be
computed independently \cite{gauge,abel}.
Since the resulting running for $e^2$
is only logarithmic in $d=4$, it is a good approximation
to neglect it in the following and use a constant $e^2$.
The ``threshold'' functions $L^4_1$
are given by eq.~(\ref{six}).
The contribution of the radial mode introduces higher
derivatives in the evolution equation, making an explicit
solution impossible. We shall again neglect this contribution,
as in the first part of the paper.
We should point out, however, that the resulting approximate
evolution equation does not become exact
any more in the large $N$ limit.
The purpose of this section is simply to demonstrate that
the correct physical behaviour is incorporated in the
full evolution equation, even though the approximate
equation that we are using does not permit quantitative accuracy.
After the omission of the contribution of the radial mode,
the resulting partial differential equation is
first order and can be solved with the
method of characteristics. We have not managed to obtain an
analytical solution in closed form, even though a numerical solution
is possible. For this reason we make an additional
approximation which is not crucial for the physical behaviour that
we are interested in (see below). We set $L^4_1(u')= L^4_1(0) = -2$
in the contributions of the Goldstone modes, while
maintaining the full ``threshold'' function in the contribution
of the gauge field. As a result we cannot observe the
decoupling of the scalar modes or the approach to convexity for the
effective potential. However, we preserve the full effect of
the gauge field on the form of the potential.
We thus finally arrive to the following evolution equation
\be
\frac{\partial u'}{\partial t}
- 2 \rht \frac{\partial u'}{ \partial \rht}
+ B \frac{\partial u'}{ \partial \rht}
+ \frac{D e^2}{1 + 2 e^2 \rht}
+ 2 u' = 0,
\label{eightyone} \ee
with
\beq
B =& 2 (N-1) v_4 = \frac{N-1}{16 \pi^2}
\nonumber \\
D =& 12  v_4 = \frac{3}{8 \pi^2}.
\label{extrathree} \eeq

The most general solution of eq.~(\ref{eightyone}) is given
by
\be
\frac{u'}{2 \rht -B}
+ \frac{D e^2}{2 (B e^2 +1)} \frac{1}{2 \rht - B}
- \frac{D e^4}{2 (B e^2 +1)^2}
\ln\left( \frac{2e^2 \rht +1}{|2 \rht - B|} \right)
= F\left( (2 \rht - B) e^{2t} \right).
\label{eightytwo} \ee
The function $F$ is determined through the initial
condition for the potential. Assuming a quartic classical
potential given by eqs.\ (\ref{fourteen}), (\ref{fourthree}) we
find
\be
F(x) =
\lx_\Lx \frac{x+B}{2 x} -\lx_\Lx \kappa_\Lx \frac{1}{x}
+ \frac{D e^2}{2 (B e^2 +1)} \frac{1}{x}
- \frac{D e^4}{2(B e^2 +1)^2}
\ln\left( \frac{e^2(x+B)+1}{|x|} \right).
\label{eightythree} \ee
In fig.~5 we plot the potential which results from eqs.
(\ref{eightytwo}), (\ref{eightythree}) for a certain
choice of the parameters of the theory.
Initially the effective average potential has only
one minimum at a non-zero value of $\rho$.
As $k$ is lowered a second minimum appears around
zero, which eventually becomes the absolute minimum
of the potential. The discontinuity in the
expectation value signals the presence of
a first order phase transition.
The development of the minimum around zero is caused by
the logarithmic terms in eqs.\
(\ref{eightytwo}), (\ref{eightythree}).
The situation is typical of a Coleman-Weinberg
phase transition triggered by radiative corrections
\cite{colwein}.
The effective potential $U = U_0$
can be calculated from eqs.\
(\ref{eightytwo}), (\ref{eightythree})
in the limit
$t \rightarrow -\infty$, with $u',\rht \rightarrow \infty$,
so that
$u'e^{2t} \sim 1$, $\rht e^{2t} \sim 1$.
We find
\be
\frac{U'(\rho)}{\Lx^2}=
\lx_\Lx \left[ \frac{\rho}{\Lx^2} -
\left( \kx_\Lx
- \frac{B}{2}
- \frac{D e^2}{2 \lx_\Lx (B e^2 +1)}
\right) \right]
+ \frac{D e^4}{(B e^2 +1)^2}
\frac{\rho}{\Lx^2} \ln\left( \frac{2 e^2
\frac{\rho}{\Lx^2}}{e^2
\left( 2 \frac{\rho}{\Lx^2}+B \right) +1} \right).
\label{eightyfour} \ee
Without the logarithmic term the phase transition
in dependence to $\kx_\Lx$ would have been second order.
The presence of the last term in eq.~(\ref{eightyfour})
results in the development of a
barrier near a secondary minimum at the
origin. This leads to a weakly first order transition, with
a discontinuity for the expectation value much smaller than
the minimum of the classical potential.
(For a detailed discussion of
the Coleman-Weinberg transition using
the full evolution equation see ref.\ \cite{abel}.)
The effective potential of eq.~(\ref{eightyfour})
is not convex. As we have mentioned already, the reason
for this is the
approximation
of the ``threshold'' function by a constant in the evolution equation.
Starting from eq. (\ref{eightyfour}) we can derive a relation
between the different mass scales in this model. We denote the
scalar field mass at the minimum $\rho_0$ of the potential by
$m^2=2 U''(\rho_0)\rho_0$, the  gauge field mass by $M^2=2 e^2 \rho_0$,
and the scalar mass around the origin
by $m_s^2= U'(0)$.
We find
\be
\frac{m^2}{M^2}+2\frac{m_s^2}{M^2}=
\frac{3e^2}{8\pi^2}\frac{1}{(Be^2+1)(Be^2+1+M^2/\Lambda^2)}.
\label{rel} \ee
The well-known result of Coleman and Weinberg \cite{colwein}
is obtained for $M^2/\Lx \ll 1$, if we
neglect the scalar contributions by setting
$B=0$, and
consider a potential with zero curvature at the
origin by setting $m_s^2=0$.

\setcounter{equation}{0}
\renewcommand{\theequation}{{\bf 4.}\arabic{equation}}

\section*{4.
Conclusions}

We presented analytical solutions of the
exact renormalization group equation for the effective average
action which are not restricted in the
vicinity of a possible infrared attractive fixed point.
This allowed us to investigate universal and non-universal
aspects of phase transitions for a variety of models.
We neglected the effects of wavefunction renormalization
and approximated the action by the potential and
a standard kinetic term.
We solved the evolution equation for the potential as a function
of the field and the running scale $k$.
We presented analytical solutions for the
$O(N)$-symmetric scalar theory in the large $N$ limit
in three, four and two dimensions.
The omission of the effects of wavefunction renormalization is
justified in four and three dimensions
by the vanishing of the anomalous dimension
to leading order in $1/N$. Parts of our results
have been obtained in the past \cite{vv,review}. However, the
emphasis of our work lies in the detailed study of the
influence of the classical
parameters of the theory on the possible phase transitions.
We also gave a crude treatment of
the Abelian Higgs model in four dimensions.
The enhanced complexity of the evolution equation for this last model
made necessary the use of additional approximations, which, however,
do not affect the qualitative conclusions.

a) For the $O(N)$-symmetric scalar theory in the large $N$ limit
in three dimensions we
distinguish two types of behaviour:
I) For a classical $\phi^4$
potential given by eq.~(\ref{fourteen}) the renormalized
theory has a second order phase transition in dependence on
$\kx_\Lx$. The universal behaviour near the transition
is governed by the Wilson-Fisher fixed point and can be parametrized
by critical exponents. II) For a classical $\phi^6$ potential
given by eq.~(\ref{fifty})
there is a parameter range for which the renormalized theory has
a second order phase transition in dependence on $\kx_\Lx$, with
universal critical behaviour. For another parameter range the
theory has a first order phase transition. The two regions are
separated by a tricritical line (at $\lx_\Lx$ = 0) which displays
tricritical behaviour with mean field exponents. \\
b) For the same theory in four dimensions the solution
reproduces the ``triviality'' of the critical theory
and the vanishing of the renormalized quartic coupling. \\
c) In two dimensions the solution demonstrates that the renormalized
theory cannot be defined in the phase with spontaneous symmetry
breaking and there is no phase transition, in agreement with the
Mermin-Wagner theorem. \\
d) For the Abelian Higgs model in four dimensions we reproduce the
Coleman-Weinberg first order phase transition
which is triggered by radiative corrections.

Our results on the universal behaviour of the three-dimensional
scalar theory and on the four-dimensional
Abelian Higgs model are in agreement with
refs. \cite{trans,indices,num,abel}, where the evolution equations
for the effective average action were studied with other methods.
Thus they provide an additional
argument for the validity and applicability of the method
of the effective average action in a wide range of problems.
The most important aspect of our solutions, however, is that
they are fully analytical and not restricted in the vicinity of
a possible infrared fixed point.
They provide a concise,
transparent picture of universal and non-universal
behaviour, at all values of the field and
the effective infrared cutoff, for a variety of physical systems.

\vspace{0.3cm}
\noindent
{\bf Acknowledgements:}
We would like to thank C. Wetterich for useful discussions
and comments. D.F.L. thanks the
Department of Theoretical Physics of the
University of Oxford for hospitality
during the course of the work.

\newpage

\section*{Figure captions}

\renewcommand{\labelenumi}{Fig.~\arabic{enumi}}
\begin{enumerate}
\item  %Fig.~1 \\
The effective average potential $U_k(\rho)$ at different scales
for a scalar model in the large $N$ limit. The
classical potential is given by eq.~(\ref{fourteen})
with $\kappa_\Lx = 1.2 NC$ and $\lx_\Lx = 0.3/NC$ $(d=3)$.
\item  %Fig.~2 \\
The derivative $u'(\rht)$ of the rescaled potential at different scales
for a scalar model in the large $N$ limit. The
classical potential is given by eq.~(\ref{fourteen})
with $\kappa_\Lx$ slightly smaller than $NC$
and $\lx_\Lx = 0.3/NC$ $(d=3)$.
\item  %Fig.~3 \\
The derivative $u'(\rht)$ of the rescaled potential
at different scales
for a scalar model in the large $N$ limit. The
classical potential is given by eq.~(\ref{fifty})
with $\kappa_\Lx = NC$,
$\lx_\Lx = 0.3/NC$ and $\sx_\Lx = 0.84/(NC)^2$
$(d=3)$.
\item  %Fig.~4 \\
The effective average potential $U_k(\rho)$ at different scales
for a scalar model in the large $N$ limit. The
classical potential is given by eq.~(\ref{fifty})
with $\kappa_\Lx = 1.2 NC$, $\lx_\Lx = 0.3/NC$
and  $\sx_\Lx = 1/(NC)^2$ $(d=3)$.
\item  %Fig.~5 \\
The approximate
effective average potential $U_k(\rho)$ at different scales
for an Abelian Higgs model with $e^2 = 2$ and $N=11$ real scalars.
The classical potential is given by eq.~(\ref{fourteen})
with $\kappa_\Lx = 1$ and $\lx_\Lx = 0.01$
$(d=4)$.
\end{enumerate}


\newpage
\begin{thebibliography}{99}


\bibitem{reneq1}
K.G. Wilson and I.G. Kogut, Phys. Rep. {\bf 12}, 75 (1974).

\bibitem{reneqex}
F.J. Wegner and A. Houghton, Phys. Rev. A {\bf 8}, 401 (1973).

\bibitem{reneq2}
F.J. Wegner, in: Phase Transitions and Critical Phenomena, vol. 6,
eds. C. Domb and M.S. Greene, Academic Press (1976);
S. Weinberg, Critical phenomena for field theorists, in Erice
Subnuc. Phys. 1 (1976).

\bibitem{nicoll}
J.F. Nicoll and T.S. Chang, Phys. Lett. A {\bf 62}, 287 (1977);
T.S. Chang, J.F. Nicoll and J.E. Young, Phys. Lett. A {\bf 67}, 287 (1978).

\bibitem{polchinski}
J. Polchinski, Nucl. Phys. B {\bf 231}, 269 (1984).

\bibitem{hasenfratz}
A. Hasenfratz and P. Hasenfratz, Nucl. Phys. B {\bf 270},
687 (1986).

\bibitem{review}
T.S. Chang, D.D. Vvedensky and J.F. Nicoll, Phys. Rep. {\bf 217},
279 (1992).

\bibitem{exeq}
C. Wetterich, Phys. Lett. B {\bf 301}, 90 (1993).

\bibitem{gauge}
M. Reuter and C. Wetterich,
Nucl. Phys. B {\bf 391}, 147 (1993);
{\it ibid} {\bf 408}, 91 (1993);
{\it ibid} {\bf 417}, 181 (1994);
{\it ibid} {\bf 427}, 291 (1994).

\bibitem{ellwanger}
U. Ellwanger, Phys. Lett. B {\bf 335}, 364 (1994).

\bibitem{pertren}
G. Keller and C. Kopper, Phys. Lett. B {\bf 273}, 323 (1991);
M. Bonini, M. D'Attanasio and G. Marchesini,
Nucl. Phys. B {\bf 409}, 441 (1993);
{\it ibid} {\bf 418}, 81 (1994);
{\it ibid} {\bf 421}, 429 (1994);
R.D. Ball and R.S. Thorne, Ann. Phys. {\bf 236}, 117 (1994).

\bibitem{epsilon}
K. Wilson and M. Fisher, Phys. Rev. Lett. {\bf 28}, 240 (1978).

\bibitem{bound}
U. Ellwanger, Z. Phys. C {\bf 58}, 619 (1993);
U. Ellwanger and C. Wetterich,
Nucl. Phys. B {\bf 423}, 137 (1994).

\bibitem{vv}
D.D. Vvedensky, J. Phys. A {\bf 17}, L251 (1984);
{\it ibid} {\bf 20}, L197 (1987).

\bibitem{averact}
C. Wetterich, Nucl. Phys. B {\bf 352}, 529 (1991); Z. Phys. C
{\bf 57}, 451 (1993).

\bibitem{trans}
N. Tetradis and C. Wetterich, Nucl. Phys. B {\bf 398}, 659 (1993);
Int. J. Mod. Phys. A {\bf 9}, 4029 (1994).

\bibitem{indices}
N. Tetradis and C. Wetterich,
Nucl Phys. B {\bf 422}, 541 (1994).

\bibitem{twoscalar}
S. Bornholdt, N. Tetradis and C. Wetterich,
Phys. Lett. B {\bf 348}, 89 (1995);
preprint HD-THEP-94-28 and OUTP-95-02 P.

\bibitem{abel}
D.F.~Litim, N. Tetradis and C. Wetterich,
preprint HD-THEP-94-23 and OUTP-94-12 P.

\bibitem{num}
J. Adams, J. Berges, S. Bornholdt, F. Freire, N. Tetradis and
C. Wetterich, preprint
CAU-THP-95-10, HD-THEP-95-15 and OUTP 95-12 P;
J. Berges, N. Tetradis and C. Wetterich, preprint
HD-THEP-95-27 and OUTP 95-27 P.

\bibitem{numer}
T.R. Morris, Phys. Lett. B {\bf 329}, 241 (1994);
M. Alford, Phys. Lett. B {\bf 336}, 237 (1994).

\bibitem{morris}
T.R. Morris, Phys. Lett. B {\bf 334}, 355 (1994).

\bibitem{spherical}
T.H. Berlin and M. Kac, Phys. Rev. {\bf 86}, 821 (1952);
H.E. Stanley, Phys. Rev. {\bf 176}, 718 (1968);
S.-K. Ma, Rev. Mod. Phys. {\bf 45}, 589 (1973);
J. Math. Phys. {\bf 15}, 1866 (1974).

\bibitem{zinn}
J. Zinn-Justin, Quantum field theory and critical phenomena, Oxford Science
Publications (1989).

\bibitem{largen}
M. Reuter, N. Tetradis and C. Wetterich,
Nucl. Phys. B {\bf 401}, 567 (1993).

\bibitem{david}
F. David, D.A. Kessler and H. Neuberger, Nucl. Phys. B
{\bf 257}, 695 (1985).

\bibitem{bmb}
W.A. Bardeen, M. Moshe and M. Bander, Phys. Rev. Lett..
{\bf 52}, 1188 (1983).

\bibitem{convex}
A. Ringwald and C. Wetterich, Nucl. Phys. B {\bf 334}, 506 (1990);
N. Tetradis and C. Wetterich, Nucl. Phys. B {\bf 383}, 197 (1992).

\bibitem{mermin}
N.D. Mermin and H. Wagner, Phys. Rev. Lett. {\bf 17}, 1133 (1966).

\bibitem{colwein} S. Coleman and E. Weinberg, Phys. Rev.
D {\bf 7}, 1888 (1973).





\end{thebibliography}
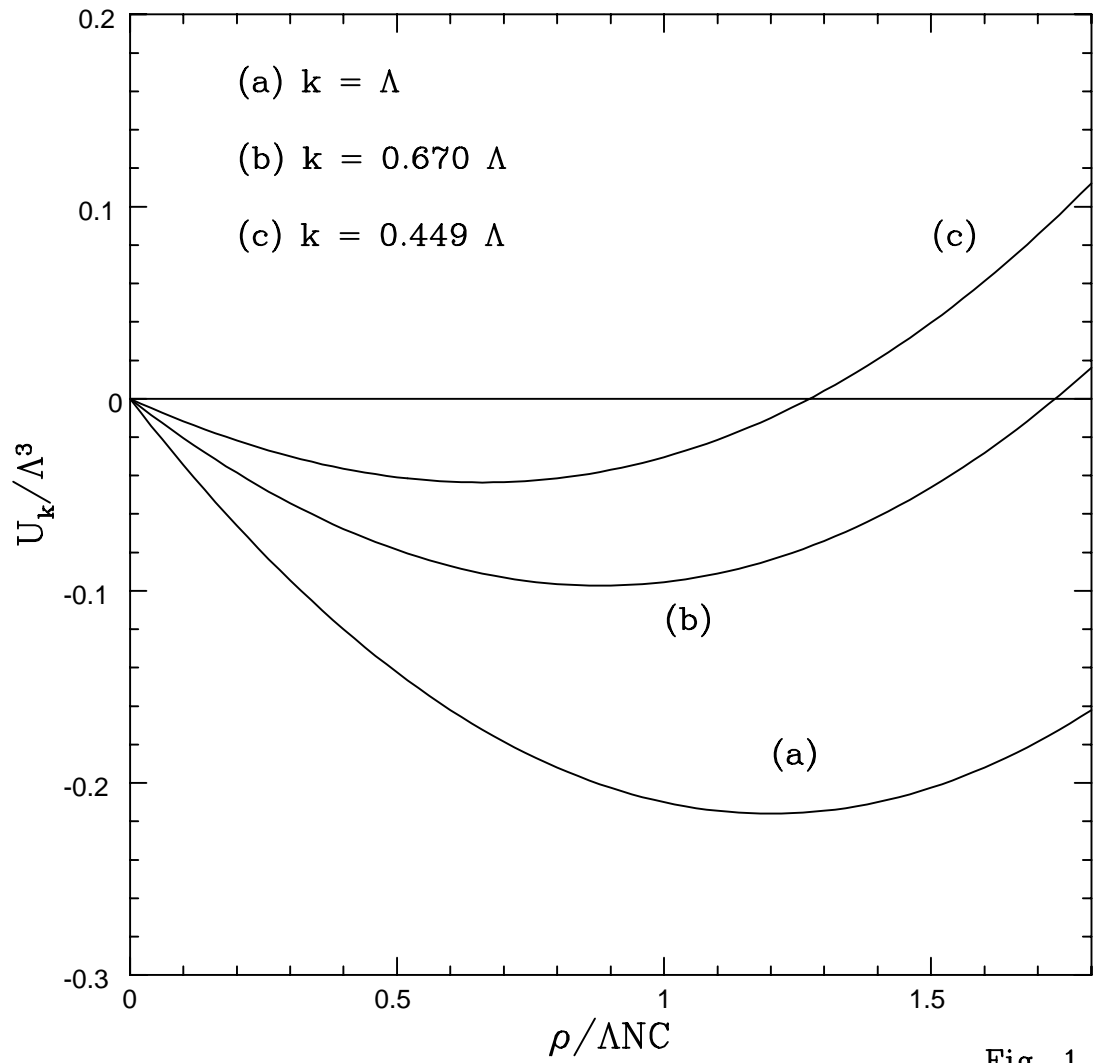
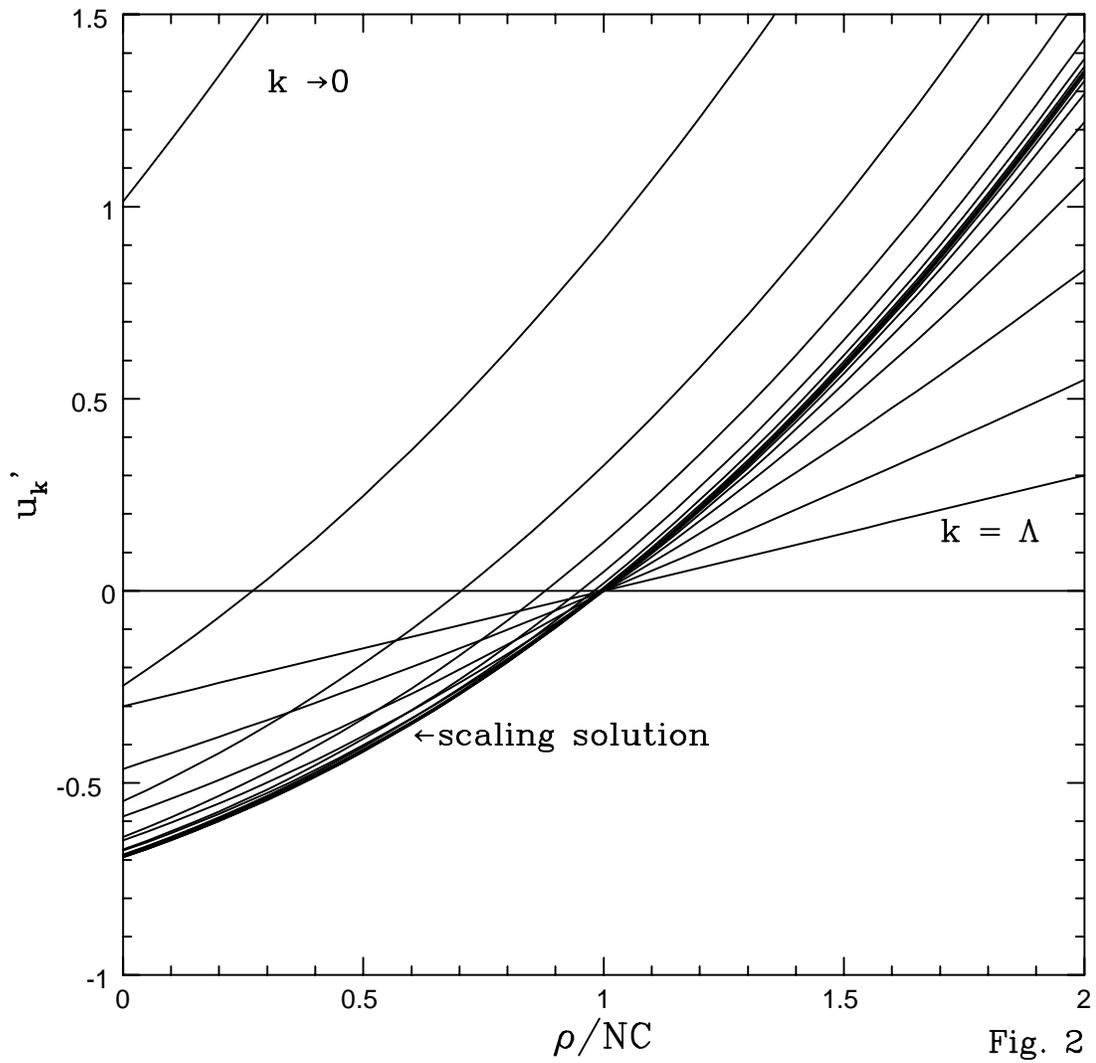
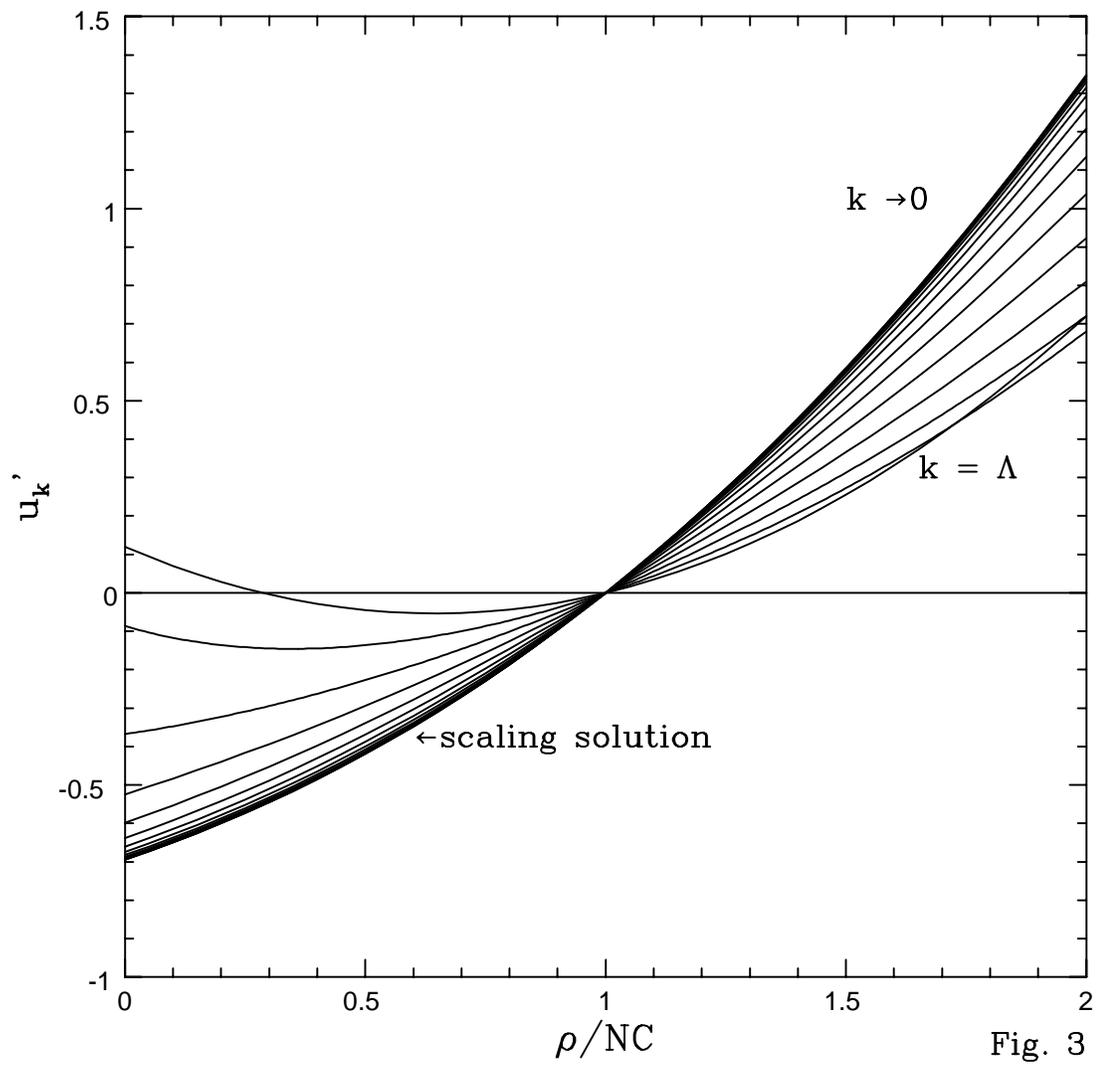
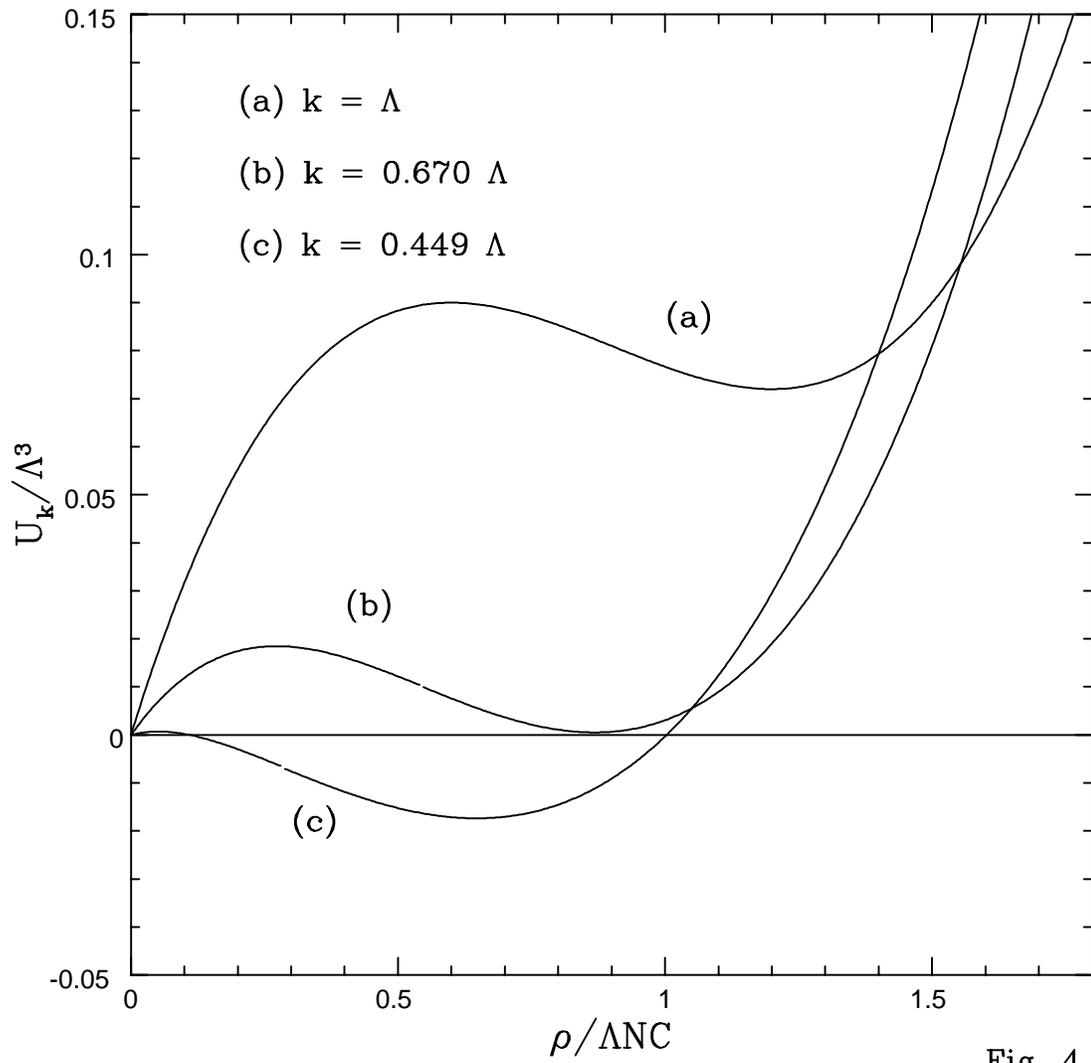
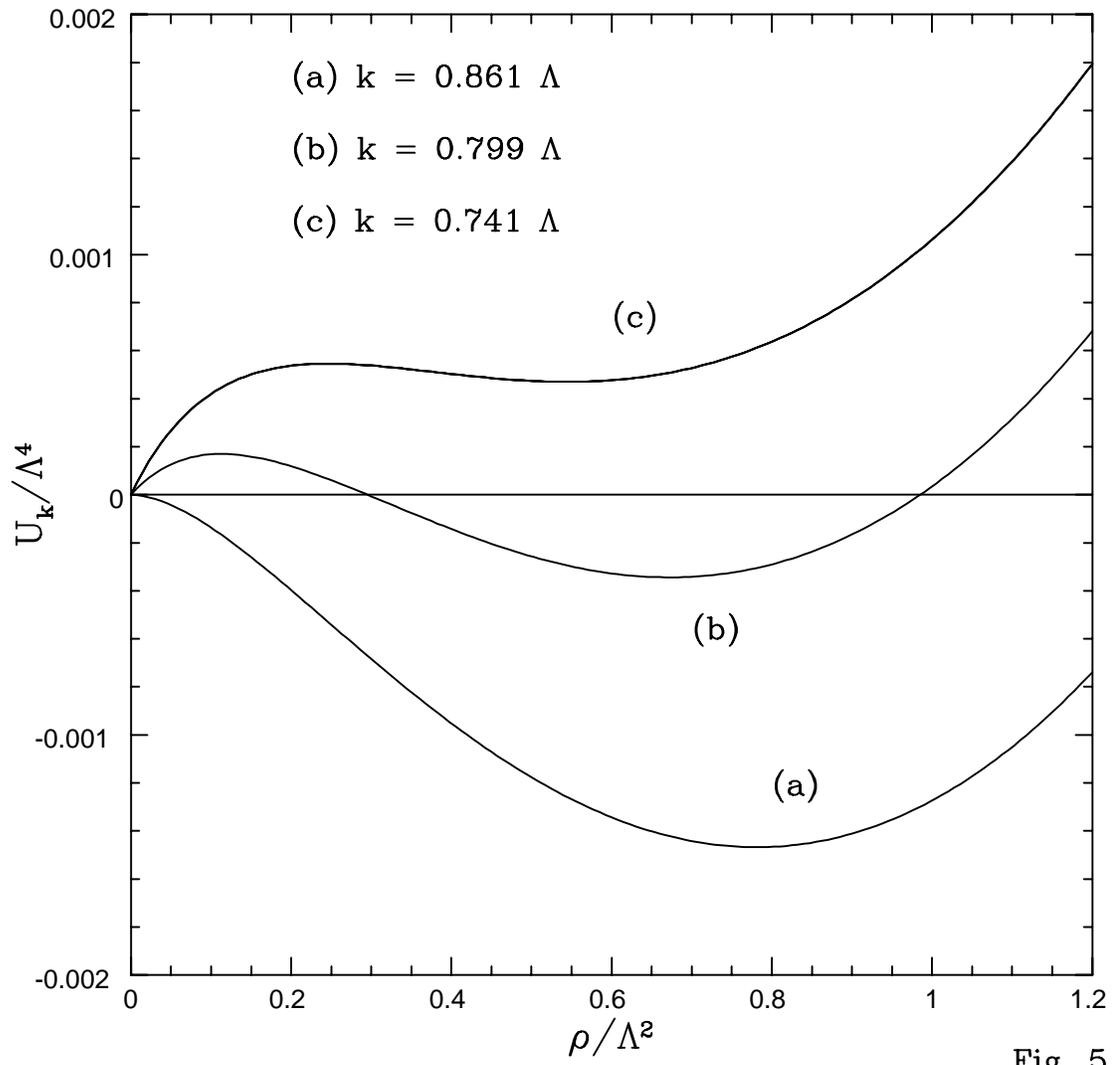
\end{document}